\documentclass{elsart}
\usepackage{amssymb}
\usepackage{graphicx}

\def\d{{\rm d}}
\def\dk{{\rm d}k}
\def\dt{{\rm d}t}

\def\x{{\bf x}}
\def\k{{\bf k}}
\def\p{{\bf p}}
\def\u{{\bf u}}
\begin{document}

\journal{Physica D}
%\pubyear{}
%\volume{}

\begin{frontmatter}

\title{Diminishing inverse transfer and non-cascading dynamics 
in surface quasi-geostrophic turbulence}

\author{Chuong V. Tran}

\address{Mathematics Institute, University of Warwick, Coventry CV4 7AL, UK}

\ead{ctran@maths.warwick.ac.uk}

\begin{abstract}
The inverse transfer in two-dimensional turbulence governed by the 
surface quasi-geostrophic (SQG) equation is studied. The nonlinear 
transfer of this system conserves the two quadratic quantities 
$\Psi_1=\langle|(-\Delta)^{1/4}\psi|^2\rangle/2$ and 
$\Psi_2=\langle|(-\Delta)^{1/2}\psi|^2\rangle/2$ (kinetic energy), 
where $\psi$ is the streamfunction and $\langle\cdot\rangle$ denotes 
a spatial average. In the limit of infinite domain, the kinetic energy 
density $\Psi_2$ remains bounded. For power-law inverse-transfer 
region, the inverse flux of $\Psi_1$ diminishes as it proceeds toward 
sufficiently low wavenumbers, implying that no persistent inverse 
cascade of $\Psi_1$ is sustainable. The unrealizability of an inverse 
cascade of $\Psi_1$ implies that there is no direct cascade of $\Psi_2$. 
Hence, the dual-cascade picture which is widely believed to be 
realizable in two-dimensional Navier--Stokes turbulence does not 
apply to SQG turbulence. Numerical results supporting the theoretical 
predictions are presented. 
\end{abstract}

\begin{keyword}

Surface quasi-geostrophic turbulence \sep Inverse transfer 
\sep Diminishing inverse flux 

\PACS 47.27.Ak \sep 47.52.+j \sep 47.27.Gs

\end{keyword}

\end{frontmatter}

\section{Introduction}

The motion of a three-dimensional stratified rotating fluid is 
characterized by the geostrophic balance between the Coriolis force 
and pressure gradient. The dynamics governed by the first order 
departure from this linear balance is known as quasi-geostrophic
dynamics (see for example \cite{Charney48,Charney71,Pedlosky87,Rhines79}),
which can be described in terms of the (three-dimensional) geostrophic 
streamfunction $\psi(\x,t)$. The vertical dimension $z$ is usually 
taken to be semi-infinite and doubly periodic conditions are usually
imposed on the horizontal flow. Normally decay conditions are required 
as $z\rightarrow\infty$. At the flat surface boundary $z=0$, the 
vertical gradient of $\psi(\x,t)$ matches the temperature field 
$T(\x,t)$, i.e. $T(\x,t)|_{z=0}=\partial_z\psi(\x,t)|_{z=0}$. 
For flows with zero potential vorticity, this surface temperature 
field can be identified with $(-\Delta)^{1/2}\psi$, where $\Delta$ is 
the (horizontal) two-dimensional Laplacian. The conservation equation 
governing the advection of the active temperature field 
$(-\Delta)^{1/2}\psi$ by the surface flow 
$\u=(-\partial_y\psi,\partial_x\psi)$
is \cite{Blumen78,Held95,Pedlosky87,Pierrehumbert94} 
\begin{eqnarray}
\label{Tadvection}
\partial_t(-\Delta)^{1/2}\psi+J(\psi,(-\Delta)^{1/2}\psi)&=&0,
\end{eqnarray}
where $J(\vartheta,\theta)=\partial_x\vartheta\,\partial_y\theta
-\partial_x\theta\,\partial_y\vartheta$.
Eq. (\ref{Tadvection}) is known as the SQG equation.

In this paper, a forced-dissipative version of (\ref{Tadvection}) 
is studied. A dissipative term of the form $\mu\Delta\psi$, where 
$\mu>0$, which results from Ekman pumping at the surface, is 
considered (cf. \cite{Constantin02,T04}). Since $(-\Delta)^{1/2}\psi$ 
is the advected quantity, this physical dissipation mechanism 
corresponds to the hypoviscous dissipation operator 
$\mu(-\Delta)^{1/2}$. The dissipation coefficient $\mu$ has the 
dimensions of velocity and is not vanishingly small in the 
atmospheric context. The system is assumed to be driven by a forcing 
$f$. Thus, the forced-dissipative SQG equation can be written as 
\begin{eqnarray}
\label{governing}
\partial_t(-\Delta)^{1/2}\psi+J(\psi,(-\Delta)^{1/2}\psi)
&=&\mu\Delta\psi+f.
\end{eqnarray}

The Jacobian operator $J(\cdot,\cdot)$ admits the identities
\begin{eqnarray}
\label{id}
\langle\phi J(\vartheta,\theta)\rangle
=-\langle\vartheta J(\phi,\theta)\rangle
=-\langle\theta J(\vartheta,\phi)\rangle,
\end{eqnarray}
where $\langle\cdot\rangle$ denotes the spatial average. As a 
consequence, the nonlinear term in (\ref{governing}) obeys the 
conservation laws
\begin{eqnarray}
\label{conservation}
\langle\psi J(\psi,(-\Delta)^{1/2}\psi)\rangle=
\langle(-\Delta)^{1/2}\psi J(\psi,(-\Delta)^{1/2}\psi)\rangle=0.
\end{eqnarray}
It follows that the two quadratic quantities 
$\Psi_1=\langle|(-\Delta)^{1/4}\psi|^2\rangle/2$ and
$\Psi_2=\langle|(-\Delta)^{1/2}\psi|^2\rangle/2$ 
(kinetic energy) are conserved by nonlinear transfer. 

The simultaneous conservation of $\Psi_1$ and $\Psi_2$ by advective 
nonlinearities imposes strict constraints on the transfer 
(redistribution) of these quantities in wavenumber space. Let us 
consider the transfer of an amount $\Psi_1=\epsilon$, initially
distributed around a given wavenumber $s$, which corresponds to an 
initial $\Psi_2=s\epsilon$. Let $\Psi_1(k)$ and $\Psi_2(k)$ be the 
resulting redistributions of $\Psi_1=\epsilon$ and of $\Psi_2=s\epsilon$. 
Given arbitrary wavenumbers $p<s$ and $q>s$, one has
\begin{eqnarray}
\label{transfer1}
\frac{1}{\epsilon}\int_q^\infty\Psi_1(k)\,\dk &\le&
\frac{1}{q\epsilon}\int_q^\infty\Psi_2(k)\,\dk \le \frac{s}{q},\\
\label{transfer2}
\frac{1}{s\epsilon}\int_0^p\Psi_2(k)\,\dk &\le&
\frac{p}{s\epsilon}\int_0^p\Psi_1(k)\,\dk \le \frac{p}{s},
\end{eqnarray}
where the two conservation laws and straightforward inequalities have
been used. The left-hand side of (\ref{transfer1}) [(\ref{transfer2})] 
is the fraction of $\epsilon$ [$s\epsilon$] that gets transferred to 
wavenumbers $k\ge q$ [$k\le p$]. This fraction is bounded from above 
by $s/q$ [$p/s$], implying that no significant fraction of $\epsilon$ 
[$s\epsilon$] can be transferred to wavenumbers $k\gg s$ [$k\ll s$]. 
This type of constraint on the nonlinear transfer of the invariants is 
a common feature in incompressible fluid systems in two dimensions. 
(Some familiar systems in this category are the Charney--Hasegawa--Mima 
equation \cite{Hasegawa78,Hasegawa79} and the class of $\alpha$ turbulence 
equations \cite{Pierrehumbert94}, which includes both the Navier--Stokes 
and the SQG equations.) The implication is that when the said initial 
sources spread out in wavenumber space, $\Psi_1$ [$\Psi_2$] is 
preferentially transferred toward lower [higher] wavenumbers. According 
to the classical theory of two-dimensional turbulence 
\cite{Batchelor69,Kraichnan67,Kraichnan71,Leith68}, which was originally 
formulated for high-Reynolds number Navier--Stokes fluids and 
subsequently thought to apply to two-dimensional incompressible fluids 
in general, this preferential transfer achieves the extreme limit by 
transferring virtually all $\epsilon$ to $k\ll s$ (inverse 
cascade) and virtually all $s\epsilon$ to $k\gg s$ (direct cascade). 
The transfer of the invariants in this manner is known as the dual 
cascade.\footnote{For some recent discussion on the possibility of a 
dual cascade in various two-dimensional systems, including the 
Navier--Stokes and SQG equations, see \cite{T04,TB03a,TB04,TS02}.} 
  
However, it is shown in this work that the preferential transfer of 
$\Psi_1$ and of $\Psi_2$ is not as dramatic as predicted but rather has 
a limited extent. More accurately, it is shown that an upper bound on 
the inverse flux of $\Psi_1$ across a low wavenumber $\ell$ vanishes 
(uniformly in time) as $\ell/s\rightarrow0$, where $s$ is the 
characteristic forcing wavenumber, thereby ruling out the existence of 
a persistent inverse cascade of $\Psi_1$ toward the low wavenumbers. 
The unrealizability of an inverse cascade implies that there is no 
direct cascade of $\Psi_2$ \cite{T04}. The physical reasons behind this 
behaviour are that the kinetic energy density of SQG dynamics remains 
bounded in the limit of infinite domain, a consequence of the hypoviscous 
dissipation operator $\mu(-\Delta)^{1/2}$, and that the inverse flux of 
$\Psi_1$ across $\ell$ is proportional to the energy content of the 
wavenumber region $k\le\ell$. These two properties of SQG turbulence, 
when combined, imply that the inverse flux of $\Psi_1$ becomes smaller 
for progressively lower~$\ell$, thus ruling out the existence of a 
persistent inverse cascade.

In the next section some preliminary estimates, which are employed 
in the derivations of the main results in Section~3, are presented.
Section~4 examines the implications of the results in Section~3 for 
the long-time dynamics and spectral distribution of kinetic energy. 
Section~5 presents some numerical results in support of the theoretical
prediction of no cascades in SQG dynamics. The paper ends with some 
discussion in the final section. 

\section{Preliminary estimates}

This section presents a simple analytic inequality and reviews the
boundedness of the energy density \cite{T04}. The former applies to
bounded domains only, while the latter is valid for both bounded and
unbounded cases.  
 
For a doubly periodic domain $L\times L$, the (complex) Fourier 
representation of $\psi$ is
\begin{eqnarray}
\psi(\x)=\sum_{\k}\exp\{i\k\cdot\x\}\widehat\psi(\k).
\end{eqnarray}
Here $\k=k_0(n,m)$, where $k_0=2\pi/L$ is the lowest wavenumber and $n$ 
and $m$ integers not simultaneously zero. For a given wavenumber $\ell$, 
let $\psi^<$ and $\psi^>$ denote, respectively, the components of 
$\psi$ spectrally supported by the disk $d=\{\k|k<\ell\}$ and its 
complement $D=\{\k|k\ge \ell\}$, i.e.
\begin{eqnarray}
\psi^<=\sum_{\k\in d}\exp\{i\k\cdot\x\}\widehat\psi(\k),
~~~~\psi^>=\sum_{\k\in D}\exp\{i\k\cdot\x\}\widehat\psi(\k).
\end{eqnarray}
For the lower-wavenumber component $\psi^<$, the following 
inequality holds:
\begin{eqnarray}
\label{ineq1}
\sup_{\x}|\nabla\psi^<| &\le& \sum_{\k\in d}k|\widehat\psi(\k)| 
\le \left(\sum_{\k\in d}1\right)^{1/2}
\left(\sum_{\k\in d}k^2|\widehat\psi(\k)|^2\right)^{1/2}
= c\frac{\ell}{k_0}{\Psi^<_2}^{1/2}
\end{eqnarray}
where $c$ is an absolute constant of order unity and 
$\Psi^<_2=\langle|\nabla\psi^<|^2\rangle/2$ the large-scale
energy density associated with the wavenumbers $k<\ell$. In (\ref{ineq1}) 
the Cauchy--Schwarz inequality is used in the second step, and the sum 
$\sum_{\k\in d}1\approx \ell^2/k_0^2$ represents the number of 
wavevectors in $d$.

On multiplying (\ref{governing}) by $\psi$ and $(-\Delta)^{1/2}\psi$ 
and taking the spatial averages of the resulting equations over the 
domain, noting from the conservation laws that the nonlinear terms 
identically vanish, one obtains evolution equations for $\Psi_1$ 
and $\Psi_2$,
\begin{eqnarray}
\label{Psi1evolution}
\frac{\d}{\dt}\Psi_1&=&-2\mu\Psi_2+\langle f\psi\rangle,\\
\label{Psi2evolution}
\frac{\d}{\dt}\Psi_2&=&-2\mu\Psi_3
+\langle f(-\Delta)^{1/2}\psi\rangle,
\end{eqnarray}
where $\Psi_3=\langle|(-\Delta)^{3/4}\psi|^2\rangle/2$. Using the 
Cauchy--Schwarz and Young inequalities, one obtains the upper bounds on 
the injection terms in (\ref{Psi1evolution}) and (\ref{Psi2evolution}):
\begin{eqnarray}
\label{forcebounds}
\langle f\psi\rangle &\le& \langle|(-\Delta)^{1/2}\psi|^2\rangle^{1/2}
\langle|(-\Delta)^{-1/2}f|^2\rangle^{1/2}
\le \mu\Psi_2+\mu^{-1}F_{-2},\nonumber\\
\langle f(-\Delta)^{1/2}\psi\rangle &\le&
\langle|(-\Delta)^{3/4}\psi|^2\rangle^{1/2}
\langle|(-\Delta)^{-1/4}f|^2\rangle^{1/2}
\le \mu\Psi_3+\mu^{-1}F_{-1},
\end{eqnarray}
where the norms of $f$ are defined by
$F_{-1}=\langle|(-\Delta)^{-1/4}f|^2\rangle/2$ and 
$F_{-2}=\langle|(-\Delta)^{-1/2}f|^2\rangle/2$. These
norms are assumed to be finite. Substituting 
(\ref{forcebounds}) in (\ref{Psi1evolution}) and 
(\ref{Psi2evolution}) yields
\begin{eqnarray}
\label{evolbound1}
\frac{\d}{\dt}\Psi_1&\le&-\mu\Psi_2
+\mu^{-1}F_{-2},\\
\label{evolbound2}
\frac{\d}{\dt}\Psi_2&\le&-\mu\Psi_3
+\mu^{-1}F_{-1}.
\end{eqnarray}
Let the overline denote the asymptotic average (existence is assumed). 
One can deduce the following upper bounds for $\overline\Psi_2$ and 
$\overline\Psi_3$:
\begin{eqnarray}
\label{bound1}
\overline\Psi_2 &\le& \mu^{-2}\overline F_{-2},\\
\label{bound2}
\overline\Psi_3 &\le& \mu^{-2}\overline F_{-1}.
\end{eqnarray}  

Ineqs. (\ref{bound1}) and (\ref{bound2}) apply to both the unbounded 
and bounded domains. In the study of turbulent transfer, the spectral 
support of $f$ is required to be in the intermediate wavenumber region,
so as to render inverse- and direct-transfer ranges free of sources.
The limit of infinite domain ($k_0\rightarrow0$), for which the classical 
theory is formulated, is taken in a straightforward manner: in the limit
$L\rightarrow\infty$, the dissipation coefficient $\mu$, the injection 
densities, and the forcing characteristic length scale (cf.~\cite{TSC04})
are held fixed. In this limit both $F_{-1}$ and $F_{-2}$ are bounded if 
$F=\langle|f|^2\rangle/2$ is bounded since 
$F_{-2}\le k_{\rm m}^{-1}F_{-1}\le k_{\rm m}^{-2}F$, 
where $k_{\rm m}$ is the minimum wavenumber of the spectral support of 
$f$. A persistent inverse cascade of $\Psi_1$ toward ever-lower wavenumbers
necessarily requires $\d\Psi_1/\dt>0$, which, by (\ref{evolbound1}), 
implies $\Psi_2<\mu^{-2}F_{-2}$. Therefore, since the main concern of the 
subsequent analyses is the realizability of an inverse cascade, for the 
rest of this paper the (instantaneous) energy $\Psi_2$ is assumed to be 
bounded in the limit $k_0\rightarrow0$. 

\section{Diminishing inverse transfer}

This section reports the main result of this paper. In the subsequent
analyses, rigorous estimates are supplemented by the usual assumption 
of power-law scaling for the energy spectrum. An inverse-transfer range  
$\Psi_2(k)=ak^{-\alpha}$, for $k<s$, is assumed. Here $s$ may be 
taken to be the minimum wavenumber of the spectral support of the 
forcing. An upper bound for the total transfer of $\Psi_1$ into the
low-wavenumber region $[k_0,\ell\ll s]$, i.e. the inverse flux of 
$\Psi_1$ across $\ell$, is derived in terms of the kinetic energy 
$\Psi_2$ and of the large-scale kinetic energy $\Psi^<_2$. It is 
shown that this flux diminishes in the limit $\ell/s\rightarrow 0$. 
The implication is that no persistent inverse cascade of $\Psi_1$ is 
realizable. 

The evolution of $\Psi^<_1=\langle|(-\Delta)^{1/4}\psi^<|^2\rangle/2$ 
is governed by
\begin{eqnarray}
\label{flux1}
\frac{\d}{\dt}\Psi^<_1 
&=& -\langle\psi^< J(\psi,(-\Delta)^{1/2}\psi)\rangle
-2\mu\Psi^<_2 \nonumber\\
&=& -\langle\psi^< J(\psi^>,(-\Delta)^{1/2}\psi)\rangle
-2\mu\Psi^<_2\nonumber\\
&=&\langle(-\Delta)^{1/2}\psi J(\psi^>,\psi^<)\rangle
-2\mu\Psi^<_2 \nonumber\\
&\le&\langle|(-\Delta)^{1/2}\psi||\nabla\psi^>||\nabla\psi^<|\rangle
-2\mu\Psi^<_2,
\end{eqnarray}
where (\ref{id}) has been used in both the second and third steps. The 
final triple-product term represents an upper bound for the inverse flux 
of $\Psi_1$ across the wavenumber $\ell$ that drives the large-scale
dynamics. 

In the limit of infinite domain, no finite rigorous estimates of the
nonlinear term in (\ref{flux1}) are available. Nevertheless, since the 
norm $|\nabla\psi^<|$ represents a measure of the large-scale fluid 
velocity (associated with $k<\ell$), it could heuristically be
identified with ${\Psi^<_2}^{1/2}$. Hence, a rough estimate of 
$|\nabla\psi^<|$ would be $|\nabla\psi^<|\approx 2^{-1/2}c'\langle
|\nabla\psi^<|^2\rangle^{1/2} = c'{\Psi^<_2}^{1/2}$, where $c'$ 
is a constant. Substituting this estimate into (\ref{flux1}) yields
\begin{eqnarray}
\label{flux2}
\frac{\d}{\dt}\Psi^<_1 &\le& 2c'{\Psi^<_2}^{1/2}
\Psi_2^{1/2}{\Psi^>_2}^{1/2} -2\mu\Psi^<_2 
\le 2c'{\Psi^<_2}^{1/2}\Psi_2-2\mu\Psi^<_2,
\end{eqnarray} 
where $\Psi^>_2=\langle|\nabla\psi^>|^2\rangle/2$ is the small-scale
energy, associated with $k\ge\ell$. Since $\Psi_2$ is bounded and since
its spectrum $\Psi_2(k)$ is assumed to obey power-law scaling in the 
inverse-transfer region, $\Psi^<_2$ necessarily diminishes as 
$\ell\rightarrow0$. Therefore, the quantity $2c'{\Psi^<_2}^{1/2}\Psi_2$, 
which bounds the inverse flux of $\Psi_1$ across $\ell$, can become 
arbitrarily small for sufficiently low $\ell$. Thus, no persistent 
inverse cascade of $\Psi_1$ is realizable provided that the foregoing 
heuristic estimate of $|\nabla\psi^<|$ can be assumed.

A rigorous version of the above calculation can be deduced for the 
bounded case. By applying (\ref{ineq1}) to (\ref{flux1}) one obtains   
\begin{eqnarray}
\label{flux3}
\frac{\d}{\dt}\Psi^<_1&\le&
\sup_{\x}|\nabla\psi^<|\langle|(-\Delta)^{1/2}\psi||\nabla\psi^>|\rangle
- 2\mu\Psi^<_2 \\
&\le&
2c\frac{\ell}{k_0}{\Psi^<_2}^{1/2}\Psi_2^{1/2}{\Psi^>_2}^{1/2}
- 2\mu\Psi^<_2 \le 2c\frac{\ell}{k_0}\left(\frac{\Psi^<_2}
{\Psi_2}\right)^{1/2}\Psi_2^{3/2} - 2\mu\Psi^<_2.\nonumber
\end{eqnarray}
The difference between (\ref{flux2}) and (\ref{flux3}) is the presence
of the ratio $\ell/k_0$ in the latter. Hence, the arguments in the 
preceding paragraph go through without change if the ratio $\ell/k_0$
is either held fixed or allowed to grow at some certain rate as 
$\ell\rightarrow0$. This condition can be stated more explicitly by 
making use of the assumed spectrum $\Psi_2(k)=ak^{-\alpha}$ for $k<s$. 
In the limit $k_0\rightarrow0$, the large-scale energy $\Psi^<_2$ 
and the total energy $\Psi_2$ can be estimated as 
\begin{eqnarray}
\Psi^<_2 &=& a\int_0^\ell k^{-\alpha}\,\dk
= \frac{a}{1-\alpha}\ell^{1-\alpha}, \\
\Psi_2 &\ge& a\int_0^s k^{-\alpha}\,\dk
= \frac{a}{1-\alpha}s^{1-\alpha},
\end{eqnarray} 
where $\alpha<1$, in accord with the boundedness of energy. It follows that
\begin{eqnarray}
\label{fluxbound}
2c\frac{\ell}{k_0}\left(\frac{\Psi^<_2}{\Psi_2}\right)^{1/2}\Psi_2^{3/2}
&\le&
2c\frac{\ell}{k_0}\left(\frac{\ell}{s}\right)^{(1-\alpha)/2}\Psi_2^{3/2}.
\end{eqnarray} 
Given a bounded $\Psi_2$ and a ratio $\ell/k_0$ that does not grow as 
rapidly as $(s/\ell)^{(1-\alpha)/2}$ as $\ell$ becomes small, this 
upper bound on the inverse flux of $\Psi_1$ clearly diminishes as 
$\ell/s\rightarrow0$. Hence, for a sufficiently wide inverse-transfer 
range $[\ell,s]$, the advective nonlinearities of SQG turbulence are 
incapable of transferring a significant amount of the injection of 
$\Psi_1$ to the low-wavenumber region $[k_0,\ell]$. 

It is interesting to generalize the above calculations to other models
of incompressible fluid turbulence in two dimensions. Pierrehumbert 
et al. \cite{Pierrehumbert94} propose to consider the so-called 
$\alpha$-turbulence models, for which the unforced and inviscid dynamics
are governed by
\begin{eqnarray}
\label{alpha}
\partial_t(-\Delta)^{\alpha/2}\psi+J(\psi,(-\Delta)^{\alpha/2}\psi)&=&0,
\end{eqnarray}
where $\alpha$ is a positive number. The two invariants of this system
are $\Psi_\alpha=\langle|(-\Delta)^{\alpha/4}\psi|^2\rangle/2$ and
$\Psi_{2\alpha}=\langle|(-\Delta)^{\alpha/2}\psi|^2\rangle/2$. The
inverse transfer of $\Psi_\alpha$ across $\ell$ can be estimated as
\begin{eqnarray}
|\langle\psi^< J(\psi,(-\Delta)^{\alpha/2}\psi)\rangle|
&=&|\langle(-\Delta)^{\alpha/2}\psi J(\psi^>,\psi^<)\rangle|\nonumber\\
&\le&\langle|(-\Delta)^{\alpha/2}\psi||\nabla\psi^>||\nabla\psi^<|\rangle
\nonumber\\
&\le&\sup_{\x}|\nabla\psi^<|\langle|(-\Delta)^{\alpha/2}\psi||\nabla\psi^>|
\rangle\nonumber\\
&\le& 2c\frac{\ell}{k_0}{\Psi^<_2}^{1/2}\Psi_{2\alpha}^{1/2}{\Psi^>_2}^{1/2}.
\end{eqnarray}
There are cases for which $\Psi_2^<\rightarrow0$ in the limit 
$\ell\rightarrow0$, leading to the diminishing of the inverse flux 
across $\ell$. For example, if the direct-cascading candidate 
$\Psi_{2\alpha}$ remains bounded\footnote{This is certainly the case 
for SQG ($\alpha=1$) and Navier--Stokes ($\alpha=2$) turbulence, for 
which the energy and enstrophy, respectively, remain bounded in the 
limit of infinite domain.} in the limit $k_0\rightarrow0$ and if 
$\alpha<1$, then $\Psi_2^<\le\ell^{2-2\alpha}\Psi_{2\alpha}^<
=\ell^{2-2\alpha}\langle|(-\Delta)^{\alpha/2}\psi^<|^2\rangle/2$, 
which certainly vanishes as $\ell\rightarrow0$ because 
$\Psi_{2\alpha}^<$ ($\le\Psi_{2\alpha}$) is bounded. On the other 
hand, $\Psi_2^>$ converges toward the low wavenumbers for the same
reason. There remains the modest assumption that $\Psi_2^>$ also 
converges toward the high wavenumbers. Given all these, the preceding 
arguments of no inverse cascade go through without change. 

{\bf Remark 1.} For technical reasons, it is difficult to generalize 
the analyses in this section to the truly unbounded case (i.e. the 
case $\ell/k_0\rightarrow\infty$, regardless of the ratio $\ell/s$). 
The main difficulty lies in the intrinsic domain-size dependence of 
the nonlinear term, thereby making its rigorous estimates, such as 
the one in (\ref{flux3}) and the subsequent estimate 
(\ref{fluxbound}), diverge in the limit $k_0\rightarrow0$.

{\bf Remark 2.} If the viscous operator $\mu(-\Delta)^{1/2}$ is replaced
by one of higher degree, as is often done in numerical simulations, then 
the boundedness of energy in the limit $k_0\rightarrow0$ cannot be 
guaranteed. Nevertheless, if boundedness of energy is assumed, then the 
above calculations go through without change since they do not refer at
all to the dissipation mechanism. 

{\bf Remark 3.} The above procedure, when applied to the nonlinear term 
of the two-dimensional Navier--Stokes equations, yields an estimate of 
the inverse energy flux that is energy-dependent. This behaviour makes 
it challenging to estimate the inverse energy transfer since the energy 
is supposed to grow with progressively wider inverse-transfer range.
Nevertheless, it can be shown that for the Kolmogorov--Kraichnan $k^{-5/3}$
energy spectrum the energy that gets transferred onto $k_0$ can be bounded 
from above by a constant independent of the width of the inverse-cascading 
range, i.e. independent of $k_0$. This result, together with related issues 
such as the Kolmogorov constant and the Kraichnan conjecture of energy 
condensation at $k_0$, is the subject of a separate study.

{\bf Remark 4.} Recently, Eyink \cite{Eyink04} raised the possibility 
of energy condensation in bounded two-dimensional Navier--Stokes 
turbulence, a conjecture by Kraichnan \cite{Kraichnan67}, and argued 
that this possibility cannot be ruled out in some recent theoretical 
results. Kraichnan \cite{Kraichnan67}, in an attempt to apply his 
dual-cascade hypothesis (initially formulated for unbounded fluids) 
to bounded turbulence, predicts that the inverse energy cascade, upon 
reaching $k_0$, deposits energy to this wavenumber, and that this 
process continues until growth of energy at $k_0$ is limited by its 
own dissipation, resulting in what may be termed an ``energy condensate''. 
The energy dissipation by this condensate alone is supposed to account 
for virtually all the energy injection, so that the energy condensate 
is also an enstrophy condensate, although the latter is of a lesser 
degree. The realization of such a ``singular'' energy and enstrophy 
concentration at $k_0$ (or around $k_0$) is required to maintain the 
proposed dual cascade in the bounded case. Recent numerical results
seemed to suggest otherwise: as the turbulence approaches a steady 
state, a $k^{-3}$ energy spectrum forms at the large scales 
\cite{Borue94,TB04}. Nevertheless, the a priori exclusion of the 
Kraichnan scenario by some recent theoretical studies, such as 
Constantin et al. \cite{Constantin94}, Tran and Shepherd \cite{TS02},
and Kuksin \cite{Kuksin04}, does not seem to be fully justified. 
The SQG dynamics allows for no possibility of such a condensate.

\section{Approach to steady dynamics and spectral distribution of energy}

This section features some physical interpretations of the results
derived in the preceding section. The non-cascading dynamics of SQG 
turbulence is discussed together with a review of the constraint on 
the spectral distribution of energy derived by Tran \cite{T04}. 

It is customary in the study of 2D turbulence to consider the  
scenario in which the fluid is driven around a wavenumber $s$ by 
steady injections $\langle f\psi\rangle=\epsilon$ and 
$\langle f(-\Delta)^{1/2}\psi\rangle=s\epsilon$. The result in the 
preceding section implies that an inverse transfer of a nonzero 
fraction of $\epsilon$ to sufficiently low wavenumbers requires that
the energy spectrum in the inverse-transfer region be no shallower than 
$ak^{-1}$. But then the energy would grow at least as rapidly as 
$a\ln(s/\ell)$ as $s/\ell\rightarrow\infty$, eventually leading to a 
balance between the injection $\epsilon$ and the dissipation $2\mu\Psi_2$. 
This result suggests two plausible routes to the long-time high-Reynolds 
number dynamics (for some suitably defined Reynolds number). First, if a 
$k^{-1}$ inverse-cascading range is realized,\footnote{Incidentally,
dimensional analyses predict a $k^{-1}$ inverse-cascading range.} the 
inverse flux decreases as it proceeds toward lower wavenumbers since 
the dissipation of $\Psi_1$, given by $2\mu\Psi_2$, grows logarithmically. 
$\Psi_1$ eventually becomes steady, as described above. Second, suppose 
that an inverse-cascading range with energy spectrum shallower than 
$k^{-1}$ is realized. The inverse flux decreases for the reason discussed 
in the preceding section as it proceeds toward sufficiently low wavenumbers. 
As a result, growth of $\Psi_1$ (and of $\Psi_2$) occurs alongside the 
existing inverse-transfer range. Eventually the dissipation 
$2\mu\Psi_2$ reaches the injection $\epsilon$ and $\Psi_1$ becomes 
steady. If power-law scaling is maintained, the slope of the energy 
spectrum in the inverse-transfer range approaches $-1$. The low-wavenumber 
end of this range $\ell$ can be calculated from $2\mu a\ln(s/\ell)=\epsilon$. 

Near steady dynamics can be achieved after the inverse flux of 
$\Psi_1$ across $\ell$ becomes sufficiently less than its dissipation 
$2\mu\Psi_2$, which should then be comparable to $\epsilon$. Hence,
by replacing $\Psi_2$ in the expression on the right-hand side of
(\ref{fluxbound}) by $\epsilon/2\mu$ and requiring that the resulting
expression be no larger than $\epsilon$, one obtains a condition for
this near steady picture:
\begin{eqnarray}
\label{fluxbound1}
\left(\frac{\ell}{k_0}\right)^2\left(\frac{\ell}{s}\right)^{1-\alpha} 
&\le& \frac{\mu^3}{\epsilon},
\end{eqnarray} 
where a constant factor of order unity has been dropped. For fixed 
(per-unit-area) injection rate $\epsilon$, forcing wavenumber $s$, 
and dissipation coefficient $\mu$, no significant fraction of 
$\epsilon$ can be transferred to wavenumbers $k\le\ell$, where 
$\ell$ satisfies (\ref{fluxbound1}). $\Psi_1$ necessarily becomes 
near steady, with the low-wavenumber region $[k_0,\ell]$ at best
weakly excited. 

For steady dynamics the balances $2\mu\Psi_2=\epsilon$ and
$2\mu\Psi_3=s\epsilon$ are achieved. It follows that $s\Psi_2=\Psi_3$,
or in terms of the energy spectrum $\Psi_2(k)$,
\begin{eqnarray} 
\label{balance}
\int_{k_0}^\infty(s-k)\Psi_2(k)\,\dk = 0.
\end{eqnarray}
This equation can be used to estimate the slopes of the energy
spectrum if power-law scaling is assumed for both the inverse- and 
direct-transfer ranges. For this purpose, let us consider the 
following spectrum 
\begin{eqnarray}
\label{spectrum}
\Psi_2(k) &=& \cases{
ak^{-\alpha}&if $\ell<k<s$,\cr
bk^{-\beta}&if $s<k<k_\nu$,\cr} ~~~as^{-\alpha}=bs^{-\beta},
\end{eqnarray}
where $a,~b,~\alpha,~\beta$ are constants, and $k_\nu$ is the highest 
wavenumber in the range $k^{-\beta}$, beyond which the spectrum is 
supposed to be steeper than $k^{-\beta}$. By substituting this spectrum 
into (\ref{balance}) and making the respective substitutions $\kappa=k/s$ 
for $k<s$, and $\kappa=s/k$ for $k>s$, one obtains \cite{T04}
\begin{eqnarray}
\label{balance1}
\int_{\ell/s}^1(1-\kappa)\kappa^{-\alpha}\,\d\kappa
&\approx&\int_{s/k_\nu}^1(1-\kappa)\kappa^{\beta-3}\,\d\kappa,
\end{eqnarray} 
where the contribution from both $k\le\ell$ and $k\ge k_\nu$ has been 
dropped. It follows that if $\ell/s\ge s/k_\nu$, then $-\alpha\le\beta-3$. 
Hence, the constraint
\begin{eqnarray}
\label{slopeconstraint}
\alpha+\beta &\ge& 3
\end{eqnarray} 
holds. Since $\alpha\le1$, $\beta$ satisfies $\beta\ge2$, meaning 
that for $k>s$, the spectrum $\Psi_3(k)$ of $\Psi_3$ is no shallower 
than $k^{-1}$. Hence, the energy dissipation cannot occur mainly at 
$k\gg s$. Thus there is no direct cascade, a dynamical behaviour 
consistent with no inverse cascade of $\Psi_1$. 

{\bf Remark 5.} Although no persistent inverse flux is possible, in the 
limit $k_0\rightarrow0$, $\Psi_1$ could become unbounded, growing at a
rate that fluctuates about zero and has a vanishingly small positive
average. The unboundedness of $\Psi_1$ requires that the energy 
spectrum of the inverse-transfer range have a non-positive slope, so 
that the spectrum $\Psi_1(k)$ is at least as steep as $k^{-1}$. For 
a given set of physical parameters, the issue of whether or not 
$\Psi_1$ becomes divergent (in the limits $k_0\rightarrow0$ and
$t\rightarrow\infty$) is interesting but is beyond the scope of 
the present work. 

{\bf Remark 6.} In some sense SQG turbulence is relatively ``simple''. 
Given spectrally localized steady injections about the forcing 
wavenumber $s$, the dynamics should eventually become non-cascading.
For a $k^{-1}$ transient inverse-cascading range, the approach to 
steady dynamics is rather slow for two reasons. First, because
$\Psi_1(k)\propto k^{-2}$, the low-wavenumber end of the 
inverse-cascading range $\ell$ proceeds relatively slowly toward 
lower wavenumbers, even during the early stages for which the inverse 
cascade is relatively strong ($\d\Psi_1/dt\approx\epsilon$). Second, 
growth of energy is only logarithmic in $\ell^{-1}$, giving rise to 
rather slow growth of the dissipation of $\Psi_1$ toward the lower 
wavenumbers.

{\bf Remark 7.} Dimensional analyses, without references to any 
particular dissipation mechanisms, predict a $k^{-1}$ inverse-transfer
range and a $k^{-5/3}$ direct-transfer range. The former is consistent 
with an inverse cascade of $\Psi_1$, but the latter allows for virtually
no energy to get transferred to the small scales.\footnote{These 
predictions seem to be mutually inconsistent since the inverse-transfer of 
$\Psi_1$ via a $k^{-1}$ spectrum is incompatible with the ``frozen-in'' 
of energy due to the $k^{-5/3}$ direct-transfer range. This picture is 
quite contrary to that of two-dimensional Navier--Stokes turbulence, for 
which a $k^{-3}$ enstrophy-transfer range means that virtually all 
enstrophy gets transferred away from the forcing region, even before 
dissipative effects are considered.} Before a viscous dissipation 
mechanism is taken into consideration, a $k^{-5/3}$ spectrum means that 
virtually no energy gets transferred to the high wavenumbers. In the 
presence of a viscous dissipation operator of the form 
$\propto(-\Delta)^\delta$, for $0\le\delta\le1/3$, instead of the 
natural dissipation operator $\mu(-\Delta)^{1/2}$, a $k^{-5/3}$ 
direct-transfer range means that the spectral energy dissipation scales 
as $k^{-5/3+2\delta}$, which is no shallower than $k^{-1}$, thereby 
allowing for virtually no energy to be dissipated at its high-wavenumber 
end.   

{\bf Remark 8.} The simultaneous conservation of $\Psi_1$ and 
$\Psi_2$ leads to an increase in $\Psi_4$ (enstrophy) when an 
initial spectral peak spreads out in wavenumber space \cite{T04}. 
This explains the observed formation of strong ``fronts'' in numerical
simulations of SQG turbulence \cite{Constantin94a,Constantin94b,Majda96},
even in the absence of a direct energy cascade as discussed above. 

\section{Numerical results}

This section reports results from numerical simulations that illustrate 
the diminishing inverse transfer and no cascades of SQG dynamics.  
Numerical studies in the literature have thus far failed to recognize 
these properties of SQG turbulence (see for example 
\cite{Ohkitani97,Pierrehumbert94,Smith02,TB03b,TB05}). Tran and Bowman
\cite{TB05}, however, notice that $\Psi_1$ is ``reluctant'' to cascade 
to the large scales, even when the natural dissipation operator
$\mu(-\Delta)^{1/2}$ is replaced by ones with higher degrees, allowing 
for relatively weaker dissipation at the large scales. 

Equation (\ref{governing}) is simulated in a doubly periodic square  
of side $2\pi$, where the modal forcing $\widehat{f}(\k)$ is 
nonzero only for those wavevectors $\k$ having magnitudes lying 
in the interval $K=[9.5,10.5]$:
\begin{eqnarray}
\label{forcing}
\widehat{f}(\k)&=&\frac{s\epsilon}{N}\frac{\widehat{\psi}(\k)}
{k\sum_{|\p|=k}|\widehat{\psi}(\p)|^2},
\end{eqnarray} 
where $s\epsilon=1$ is the constant energy injection rate and~$N$ the 
number of discrete wavenumbers in~$K$. The wavenumber $s\approx10$ is 
defined such that $s^{-1}$ is the mean of $k^{-1}$ over $K$. The 
(constant) injection rate of $\Psi_1$ is $\epsilon\approx0.1$. 
Dealiased $683^2$ and $1365^2$ pseudospectral simulations ($1024^2$ 
and $2048^2$ total modes) were performed. For Navier--Stokes 
turbulence, these resolutions are sufficient to simulate an inverse 
cascade that carries about a quarter of the energy injection to the 
large scales via a discernible $k^{-5/3}$ range \cite{T04}. For SQG 
turbulence, even the higher resolution turns out to be insufficient
to simulate a noticeable transient inverse cascade. Two dissipation 
coefficients were used: $\mu=0.05$ and $\mu=0.025$. The lower and 
higher resolutions were used for the stronger and weaker dissipation, 
respectively. Both simulations were initialized with the spectrum 
$\Psi_2(k)=10^{-2}\pi k/(100+k^2)$. 

\begin{figure}
\centerline{\includegraphics{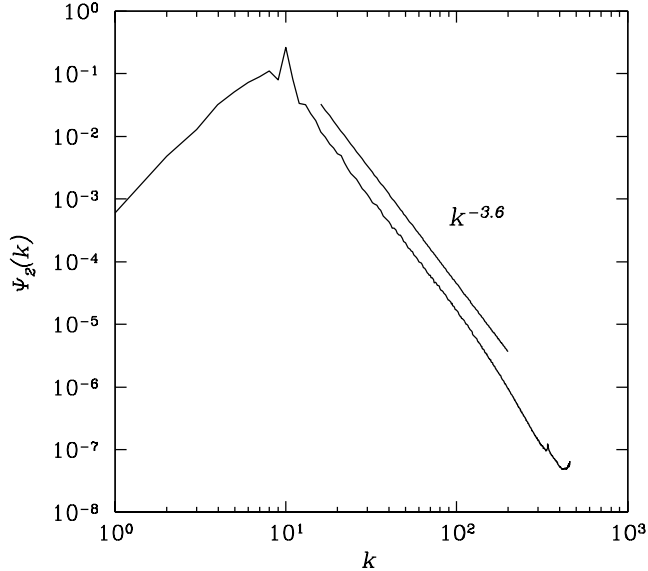}}
\caption{The time-averaged steady-state energy spectrum $\Psi_2(k)$ 
vs.~$k$ for the dissipation coefficient $\mu=0.05$. The average 
energy is $\Psi_2=0.99$, implying that the dissipation of $\Psi_1$, 
averaged in the same period, is $0.099$. This amounts to virtually all 
of the injection rate $\epsilon\approx0.1$. Hence, no inverse cascade 
of $\Psi_1$ exists and both $\Psi_1$ and $\Psi_2$ are steady.}
\label{SQG1}
\end{figure}

Figure~\ref{SQG1} shows the time-averaged (between $t=19.7$ and 
$t=20.3$)\footnote{The dissipation time at the forcing wavenumber 
$s\approx10$ is $(2\mu s)^{-1}\approx1$, so that the time $t\approx20$ 
is sufficiently long for the evolution of $\Psi_2$. In fact, it was 
observed that both $\Psi_1$ and $\Psi_2$ became steady at $t\approx 12$.} 
kinetic energy spectrum for the case $\mu=0.05$. The average energy is 
$\Psi_2=0.99$. The dissipation of $\Psi_1$, averaged for the same period, 
is $2\mu\Psi_2=0.099$, which amounts to virtually all the injection rate 
$\approx0.1$. The energy was observed to increase monotonically from 
$t=0$ to $t\approx12$, by which time $\d\Psi_1/\dt\approx0$; no 
significant inverse transfer was observed throughout the period. Instead,
growth of energy occurs mainly around the forcing region. The 
steady spectrum is relatively shallow in the low-wavenumber range (see 
Fig.~\ref{SQG1}), meaning that the lowest wavenumbers are virtually 
unexcited. The small-scale energy spectrum scales as $k^{-3.6}$, so 
that the spectrum $\Psi_3(k)$ of the energy dissipation agent $\Psi_3$
scales as $k^{-2.6}$. This scaling means that the energy dissipation 
occurs mainly around the forcing region, consistent with the weak 
inverse transfer.

\begin{figure}
\centerline{\includegraphics{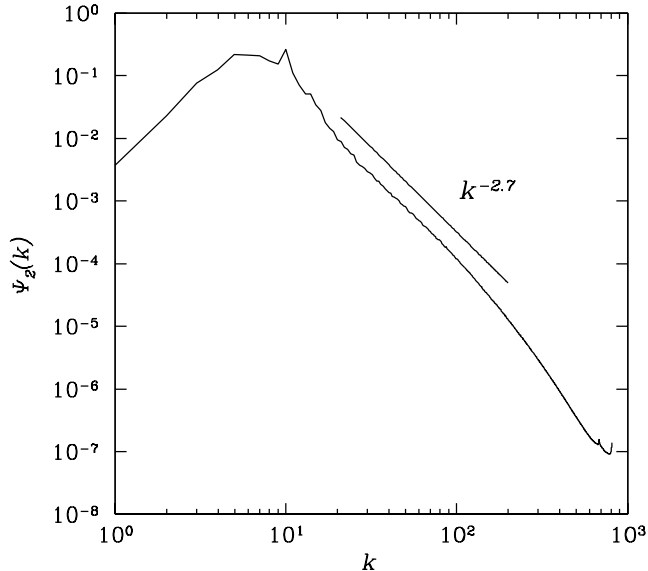}}
\caption{The time-averaged near steady-state energy spectrum $\Psi_2(k)$ 
vs. $k$ for the dissipation coefficient $\mu=0.025$. The average 
energy is $\Psi_2=1.96$, implying that the dissipation of $\Psi_1$, 
averaged in the same period, is $0.098$. This amounts to most of 
the injection rate $\epsilon\approx0.1$.}
\label{SQG3}
\end{figure}

A somewhat stronger transient inverse transfer was observed for the case 
$\mu=0.025$. Fig.~\ref{SQG3} shows a near steady kinetic energy spectrum 
averaged between $t=15.1$ and $t=15.6$. The average energy is 
$\Psi_2=1.96$. The dissipation of $\Psi_1$ averaged for the same period 
is $2\mu\Psi_2=0.098$, which amounts to virtually all of the injection 
rate $\approx0.1$. The large-scale spectrum is better ``filled up'' 
than that of the previous case, due to a stronger inverse transfer 
during the transient phase. Nevertheless, no significant fraction of 
$\epsilon$ reaches the lowest wavenumbers. The energy dissipation occurs
around the forcing region, as is evident from the steep small-scale
spectrum (see Fig.~\ref{SQG3}). 

It is expected that for higher resolutions (so that simulations with
smaller $\mu$ are possible), transient inverse fluxes can become more
noticeable and steeper inverse-transfer ranges can be realized. Given 
limited resolutions, Tran and Bowman \cite{TB05} use high-degree 
dissipation operators $\propto\Delta$ and $\propto(-\Delta)^{3/2}$. 
In both cases, weak inverse cascades are observed, but the transient 
energy spectra of the inverse-transfer region are considerably shallower 
than $k^{-1}$. These spectra cannot support a persistent inverse cascade, 
as argued in Section~3. However, an interesting possibility arises. Due
to both the inability of the nonlinear transfer to sustain a constant
(wavenumber-independent) inverse flux and $\d\Psi_1/\dt>0$, growth of 
$\Psi_1$ ought to occur within the existing inverse-transfer range, 
thereby causing this range to become steeper. This process may continue 
until the inverse-transfer range eventually exceeds the $k^{-1}$ 
threshold. Because of the high degrees of viscosity, such a slope 
steepening causes insignificant increase in the dissipation of $\Psi_1$. 
As a result, a positive growth rate $\d\Psi_1/\dt$ could be maintained 
for a steeper-than-$k^{-1}$ inverse-transfer range, which might then be 
able to support a persistent inverse cascade. 
   
\section{Conclusion}

In this paper, the advective transfer of SQG turbulence is studied.
The main result obtained is an upper bound for the inverse flux of 
the inverse-cascading candidate $\Psi_1$. This upper bound diminishes 
as the flux proceeds toward sufficiently low wavenumbers, thereby 
ruling out the existence of a persistent inverse cascade. The 
unrealizability of an inverse cascade entails that there is no direct 
cascade. This is the first rigorous example of the dynamics of 
incompressible fluids in two dimensions that exhibits no cascades.  

There are two essential features of SQG turbulence that facilitate the 
proof of 
non-cascading dynamics. First, the inverse flux of $\Psi_1$ across a 
low wavenumber $\ell$ can be uniformly (in time) estimated in terms of 
the kinetic energy $\Psi_2$ and of $\Psi_2^<$, the large-scale
energy associated with the low-wavenumber region $k<\ell$. Second, the
energy $\Psi_2$ remains bounded in the limit of infinite domain. The 
former is an intrinsic property of the advective nonlinear term, and 
the latter is due to the hypoviscous nature of the dissipation of SQG 
dynamics. The boundedness of energy means that if the turbulence is 
driven at some fixed energy density rate around some fixed wavenumber 
$s$, virtually no energy can be acquired by wavenumbers $k\ll s$. This 
means that in the limit $\ell/s\rightarrow0$, $\Psi_2^<$ can become 
arbitrarily small, leading to an arbitrarily small inverse flux of 
$\Psi_1$ across $\ell$. Hence, for sufficiently low wavenumber $\ell$ 
no significant inverse transfer of $\Psi_1$ across $\ell$ is possible.

Numerical simulations of SQG turbulence were performed. The results 
show no significant inverse transfer of $\Psi_1$ to the large scales,
thereby lending strong support to the prediction of no cascades. The 
large-scale energy spectra are shallower than $k^{-1}$, and the 
small-scale spectra are steeper than $k^{-2}$, consistent with the 
theoretical prediction.

%\bibliography{ref}

\end{document}